\documentclass[a4paper,10pt]{article}%
\usepackage{graphicx}
\usepackage[centertags]{amsmath}
\usepackage[english]{babel}
\usepackage{indentfirst}
\usepackage{amsthm}
\usepackage{newlfont}
\usepackage{amsfonts}
\usepackage{amssymb}%
\usepackage{psfrag}

\begin{document}

\author{Jean Faber$^{1},$ Renato Portugal$^{1}$ and
Luiz Pinguelli Rosa$^{2}$\\$^{1}${\small National Laboratory of
Scientific Computing - LNCC, }\\
{\small Av. Getúlio Vargas 333, Quitandinha, 25651-075, Petrópolis, RJ,
Brazil. }\\{\small \{faber, portugal\}@lncc.br}\\$^{2}${\small
Federal University of Rio de Janeiro, COPPE-UFRJ, RJ, Brazil,
}\\{\small lpr@adc.coppe.ufrj.br}}
\title{{\Large Quantum Games in Open Systems using Biophysic Hamiltonians}}
\maketitle

\begin{abstract}
We analyze the necessary physical conditions to model an open
quantum system as a quantum game. By applying the formalism of
Quantum Operations on a particular system, we use Kraus operators
as quantum strategies. The physical interpretation is a conflict
among different configurations of the environment. The resolution
of the conflict displays regimes of minimum loss of information.
\end{abstract}

\section{Introduction}

Recently, many papers have extended the game theory concepts to
the quantum world \cite{Meyer}--\cite{Chen et al}. By quantizing a
game, it is possible to improve its efficiency and stability
\cite{Benjamim1}. However, by looking at the relationship between
physical systems and the Theory of Games (TGs) we may ask: What
are the conditions to model a physical system in terms of a game?

In order to address this question we will turn our attention to
the theory of open quantum systems. Open systems allow us to use
Kraus operators as quantum strategies, which can be obtained by
the formalism of Quantum Operations \cite{Chuang}. Our goal is to
model open quantum systems using Quantum Games. We apply these
concepts to the quantum biological model proposed by Fröhlich
\cite{Frohlich1}, \cite{Pokorny}.

In section 2 we review the main concepts of quantum games. We
emphasize the use of the Kraus operators as quantum strategies. In
section 3 we discuss a biophysic Hamiltonian of an open system and
explicitly calculate the Kraus operators associated with the
coupling term of the Hamiltonian.

In section 4 we propose a game that uses the Kraus operators as
quantum strategies and a general payoff criterion for open quantum
systems. In section 5 we analyze numerically the equilibrium
points in the Quantum Phase Damping Channel game model. In section
6 we analyze the possible consequences of this new proposal.
Finally, in section 7 we draw conclusions.

\section{Extending the Quantum Strategies}

In quantum games the strategies usually belong to the group
$SU(2)$. However, to obtain relevant quantum effects in a game it
is enough to consider a 2-parameter unitary matrix defined by
\cite{Eisert1}, \cite{Benjamim1}, \cite{Abbott1}%
\begin{equation}
U(\theta,\varphi)=\left(
\begin{array}
[c]{cc}%
e^{i\varphi}\cos(\theta/2) & \sin(\theta/2)\\
-\sin(\theta/2) & e^{-i\varphi}\cos(\theta/2)
\end{array}
\right),
\end{equation}
where $\theta\in\lbrack0,\pi]$ and $\varphi$ $\in\lbrack0,\pi/2].$
Following the protocol of Eisert, Wilkens, and Lewenstein (EWL)
\cite{Eisert1}, a quantum game of two players, $A$ and $B$, can be
completely described by the set
$\Omega=\{\mathcal{H},\mathcal{S}_{A},\mathcal{S}_{B},\rho_{in},\mathcal{F}_{A},\mathcal{F}_{B}\}$,
where $\mathcal{H}$ is the Hilbert space of the quantum system,
$\rho_{in}$ is the density matrix which defines the initial
configuration of the game and $\mathcal{F}_{A,(B)}$ is the
\textit{evaluation function} or \textit{payoff function} of the
player $A$ ($B$), defined as
\begin{equation}
\mathcal{F}_{A,B}(\{\theta_{A},\varphi_{A}\},\{\theta_{B},\varphi
_{B}\})=\operatorname{tr}(P_{A,B}\rho_{\text{\textit{fin}}}),
\end{equation}
where $P_{A,B}=\sum_{ij}w_{ij}^{A,B}|ij\rangle\langle ij|$ are the
\textit{payoff operators} and
$\rho_{\text{\textit{fin}}}=|\psi_{\text{\textit{fin}}}^{AB}\rangle\langle
\psi_{\text{\textit{fin}}}^{AB}|$, with
\begin{equation}
|\psi_{\text{\textit{fin}}}^{AB}\rangle=J^{\dagger}(\chi)\left(  \frac{{}}{{}%
}U(\theta_{A},\varphi_{A})\otimes U(\theta_{B},\varphi_{B})\frac{{}}{{}%
}\right)  J(\chi)|00\rangle, \label{final state}%
\end{equation}
and $J(\chi)$ is a nonlocal unitary transformation defined by
$J(\chi)=\cos (\chi/2)I\otimes I +\sin(\chi/2)D\otimes D,$ with
$\chi\in\lbrack0,\pi/2]$ controlling the entanglement degree,
$I=U(0,0)$ and $D=U(\pi,0).$

Nevertheless, according to \cite{Eisert1} and \cite{Benjamim1}
every quantum game can be described as a \textit{Completely
Positive Trace-Preserving Map }(CPTPM). This characterization
allows an important generalization: We can use Kraus operators as
quantum strategies. In this direction, by using the formalism of
Quantum Operations (QOs) \cite{Chuang}, which defines a CPTPM, we
will be able to model an open quantum systems as a game. QOs may
describe different types of noise that different kinds of
environmental configurations introduce into a quantum system. The
noises are described by the Kraus operators which act on the
principal system.

In this way, the solutions of a game are the equilibrium and the
dominant points of the strategies. There are two important
concepts \cite{Eisert1}, \cite{Marinatto}, \cite{Hofbauer}:

\smallskip

$\mathbf{(i)}$\textit{Dominant Strategy} (DE). It is the strategy
that has the best performance in a game from the viewpoint of one
player. It doesn't necessarily give the best payoff if all players
adopt the DE.

\smallskip

$\mathbf{(ii)}$ \textit{Nash Equilibrium }(NE). It is the pair of
strategies $(\overset{\star}{s}_{A},\overset{\star}{s}_{B})$ from
which neither player can improve their payoff by an unilateral
change in strategy. Mathematically, a pair
$(\overset{\star}{s}_{A},\overset{\star}{s}_{B})$ is in NE if
\begin{subequations}
\begin{align}
\mathcal{F}_{A}(\overset{\star}{s}_{A},\overset{\star}{s}_{B})-\mathcal{F}%
_{A}(s_{A},\overset{\star}{s}_{B})  &  \geq0,\;\;\forall s_{A}, \label{Nash A}%
\\
\mathcal{F}_{B}(\overset{\star}{s}_{A},\overset{\star}{s}_{B})-\mathcal{F}%
_{B}(\overset{\star}{s}_{A},s_{B})  &  \geq0,\;\;\forall s_{B}. \label{Nash B}%
\end{align}
\end{subequations}

\section{Quantum Strategies from Hamiltonians}

In the sixties, Fröhlich \cite{Frohlich1} developed a theoretical
model focusing on some of electro-oscillatory physical properties
of biomolecular systems. He showed that it is possible to describe
relevant quantum effects in those biosystems by approximating a
group of proteins and its environment by a group of coupled
harmonic oscillators. The complete Fröhlich Hamiltonian is
described in \cite{Pokorny}. To simplify the model we consider
only the Hamiltonian that describes the interaction between the
principal system and the thermal bath
\begin{equation}%
H_{TB}=\hslash\xi(a^{\dagger}a%
b^{\dagger}+a^{\dagger}ab), \label{Hamilt Frohlich}
\end{equation}
where $a^{\dagger}$ and $a$ are the creation and the annihilation
operators of the principal system, respectively; $b^{\dagger}$ and
$b$ are the creation and annihilation operators of the thermal
bath, respectively. The Hamiltonian $H_{TB}$ describes a coupling
of \textit{second order} between the principal system and the
environment (the thermal bath), through a real coupling constant
$\xi$. Both the principal system and the environment are modeled
as harmonic oscillators.

In order to construct a game from the Hamiltonian (\ref{Hamilt
Frohlich}) we must find the superoperators associated with the
coupling term $\xi$. Each set of superoperators corresponds to an
environment noise inserted into the principal system. Therefore,
we can ask: What is the best environmental configuration that
generates the smallest loss of information in the principal
system? In other words, what is the best frequence of noise
insertion that provides the maximum storage of information in the
principal system?

From the definition of QOs, we obtain the general expression for
the superoperators \cite{Chuang}%
\begin{equation}
\mathbf{S}_{k}=\sum_{m}\sum_{n}\langle\mathfrak{A}_{m}%
\mathfrak{B}_{k}|U_{TB}|\mathfrak{A}\text{$_{n}%
\mathfrak{B}_{0}$}\rangle|\mathfrak{A}_{m}\rangle
\langle\mathfrak{A}_{n}|, \label{General Superoperator}%
\end{equation}
where $\{|\mathfrak{B}_{i}\rangle\}$ and
$\{|\mathfrak{A}_{i}\rangle\}$ are the states of the thermal bath
and the principal system respectively. The states of the
environment are taken in the thermodynamic equilibrium and the
states of the principal system are taken in a general high level
of energy. The evolution operator $U_{TB}$ obey the Schrödinger
equation associated with $H_{TB}$. Using the fact that
$[a_{i}^{\dagger},b_{j}^{\dagger}]=0,$ $\forall\;i,j$, the
relation
$U_{TB}|\mathfrak{A}_{0}\mathfrak{B}_{0}\rangle=|\mathfrak{A}_{0}\mathfrak{B}_{0}\rangle$
and the
Baker-Campbell-Hausdorf relation \cite{Chuang}, we obtain%
\begin{align}
U_{TB}|\mathfrak{A}_{n}\mathfrak{B}_{0}\rangle &
=\sum_{m=0}^{\infty}\frac{(-int\xi)^{m}}{(m!)\left(  n!\right)  ^{-1/2}%
}(b+b^{\dagger})^{m}|\mathfrak{A}_{n}\mathfrak{B}_{0}%
\rangle\\
&
=\sum_{m}^{\infty}\sum_{j=0}^{N-1}\frac{(-int\xi)^{m}}{(m!)\left(
n!\right)^{-1/2}}g(m,j)|\mathfrak{A}_{n}\mathfrak{B}_{j}\rangle,\label{Evolution
TB2}
\end{align}
where $g(m,j)$ are defined by
$(b+b^{\dagger})^{m}|0\rangle=\sum_{j=0}^{N-1}g(m,j)|j\rangle$ and
are derived from the recurrent expression

\begin{equation}
(b+b^{\dagger})^{m+1}|0\rangle=(b+b^{\dagger})(  \sum_{j=0}%
^{N-1}g(m,j)|j\rangle).
\end{equation}

For a quantum game it is enough to consider a system of only two
levels of energy. Substituting $U_{TB}$ into (\ref{General
Superoperator}) we obtain the superoperators relative to the
second order coupling between the principal system and the thermal
bath. For two levels we have to analyze the convergence of the
series with the coefficient $g(m,j)$. For all even $m$ $g(m,0)=1$,
for all odd $m$ $g(m,0)=0$ and the opposite for $g(m,1)$.

Therefore, from (\ref{Evolution TB2}), we obtain
\begin{equation}%
\begin{array}
[c]{c}%
\mathbf{S}_{0}(\gamma)=|\mathfrak{A}_{0}\rangle\langle\mathfrak{A}_{0}%
|+\cos(t\xi)|\mathfrak{A}_{1}\rangle\langle\mathfrak{A}_{1}|=\left(
\begin{array}
[c]{cc}%
1 & 0\\
0 & \sqrt{1-\gamma}%
\end{array}
\right)  ,\\
\\
\mathbf{S}_{1}(\gamma)=\sin(t\xi)|\mathfrak{A}_{1}\rangle\langle
\mathfrak{A}_{1}|=\left(
\begin{array}
[c]{cc}%
0 & 0\\
0 & \sqrt{\gamma}%
\end{array}
\right)  ,
\end{array}
\label{TB2 especificos}%
\end{equation}
where $\gamma=\sin^{2}(t\xi).$

The superoperators (\ref{TB2 especificos}) define a mapping known
as \textit{quantum} \textit{phase damping channel }(QPDC). This
channel plays an important role because it models the effect of
\textit{decoherence}. In this case $\gamma$ can be interpreted as
the probability that a photon has been scattered from a quantum
system \cite{Chuang}.

The point now is to construct a reward criterion to measure the
information in the principal system. Physically it means to
calculate directly the entropy of the principal system after
having interacted with the environment. However, we cannot say
what environment configuration provides the optimal points of
information storage. Using the TGs we will be able to analyze
these optimal points by comparing two different environmental
configurations. The principal advantage of this method is the
direct association of a physical system with an information
analysis. Besides, it provides a new interpretation of open
quantum systems as a kind of game, showing a hidden optimization
procedure in such systems.

\section{A Payoff Based on the Information of the System}

In order to construct a payoff criterion, we calculate the
information of the principal system after interaction with the
environment (for each space, $A$ and $B$). To do this, we use the
concept of information as described in Refs. \cite{Chuang} and
\cite{Pokorny}. If the superoperators represent strategies, the
rewards will be given according to the loss information of the
principal system produced by each pair of strategy (from the
players $A$ and $B$ simultaneously). If the player $A$ plays a
strategy $s_{i}^{A}\in\mathcal{S}_{A}$ and the player $B$ plays a
strategy$\;s_{k}^{B}\in\mathcal{S}_{B},$ the
joint information is given by%
\begin{equation}
I(s_{i}^{A},s_{k}^{B})=-\log_{2}\left(  p_{ik}^{AB}\right) \forall
i,k,
\label{inf conj e inf shannon}%
\end{equation}
where $p_{ik}^{AB}$ is the joint probability associated with a
quantum noise from the simultaneous occurrence of the pair
$s_{i}^{A}$ e $s_{k}^{B}$. This probability measures the degree of
dependence between the strategies of the players\footnote{To avoid
singularities, we limited the logarithm function by
$|\underset{\varepsilon\longrightarrow0}{\lim}\left(
-\log_{2}(\varepsilon )\right)  |\leq\delta,$
$\forall\;\delta\geq0.$ For numerical calculations we assume the
minimum as $\varepsilon=10^{-10}.$}. If the strategies of $A$ and
$B$ are measurement operators, the joint probabilities are given
by
\begin{equation}
p_{ik}^{AB}= \operatorname{tr} \left(  s_{i}^{A}\otimes
s_{k}^{B}\left(  \rho
_{\text{\textit{in}}}^{\prime}\right)  s_{i}^{A\dagger}{\otimes}%
s_{k}^{B\dagger}\right) , \label{prob conj}%
\end{equation}
where in the game context $
\rho_{\text{\textit{in}}}^{\prime}=J(\chi)\rho_{in}
J^{\dagger}(\chi)$ such that $\rho_{\text{\textit{in}}}=|\psi
_{\text{\textit{in}}}^{AB}\rangle\langle\psi_{\text{\textit{in}}%
}^{AB}|,\;|\psi_{\text{\textit{in}}}^{AB}\rangle
=|\mathfrak{A}_{0}^{A}\mathfrak{A}_{0}^{B}\rangle\equiv
|00\rangle$ and the set of strategies of $A$ and $B$ is
constructed from the superoperators (\ref{TB2 especificos}).

From the information theory, we know that events with small chance
to happen have high information and vice versa. Therefore, by
associating the joint information to each element of the payoff
matrix, we obtain a good criterion to reward the ``player's
choice''. Of course the environment does not choose its
strategies. The actions are physically equivalent to a random
insertion of noise into the principal system. Thus, in the
computational basis we can define the payoff operators by%
\begin{equation}
P_{A,B}= \sum_{ik}w_{ik}^{A,B}|ik\rangle\langle ik| \\
 = \sum^{1}_{i,k=0}I(s^{A}_{i},s^{B}_{k})|ik \rangle\langle ik|,\label{projectors}
\end{equation}
where $I(s_{i}^{A},s_{k}^{B})=I(s_{k}^{B},s_{i}^{A}),$ $\forall i\neq k,$
producing a symmetric game.

This criterion introduces a natural cooperation between the
players since the optimal points are reached by pairs of
strategies. Besides, this payoff criterion can be seen as a kind
of \textquotedblleft information measurement\textquotedblright\
applied to open quantum systems since the payoff function
corresponds to an average over the amount of information
associated with the quantum noise probabilities. It expresses
naturally the amount of information in the interface between the
quantum and the classical worlds. Hence, we can apply this
criterion in every quantum game that uses measurement operators as
quantum strategies\footnote{For unitary operations an alternative
criterion might be given by the \textit{Von Neumann Entropy}
\cite{Chuang}. The payoffs would be constructed from
$w^{A,B}_{ik}=-\operatorname{tr_{B,A}} \left (\phi
\operatorname{log}_{2} \phi \right ),$ where $\phi = (U^A_i
\otimes U^B_k) \rho (U^A_i \otimes U^B_k)^{\dagger}$.}.

\section{An Open System as a Quantum Game}

If we use the superoperators as quantum strategies the noise
insertion can be seen as players applying their strategies. By
considering two different environmental configurations, $A$ and
$B$, we retrieve two players. That is, the noise insertion from
each environment represents the strategic action of each player.

Since the superoperators (\ref{TB2 especificos}) satisfy the
condition
$\sum_{k}\mathbf{S}_{k}\mathbf{S}_{k}^{\dagger}=\mathbf{1}$,
we label the set of strategies of the players $A$ and $B$ by $\mathcal{S}%
_{A}=\{s_{0}^{A}\equiv\mathbf{S}_{0}(\gamma_{A}),\;s_{1}^{A}\equiv
\mathbf{S}_{1}(\gamma_{A})\}$ and
$\mathcal{S}_{B}=\{s_{0}^{B}\equiv
\mathbf{S}_{0}(\gamma_{B}),\;s_{1}^{B}\equiv\mathbf{S}_{1}(\gamma_{B})\},$
respectively. In this way, from the definition of QOs and from the
EWL protocol, the final state of the game is%
\begin{equation}
\rho_{\text{\textit{fin}}}=J^{\dagger}(\chi)\sigma J(\chi), \label{rho final}%
\end{equation}
where%
\begin{equation}
\sigma=\sum_{i,k=0}^{1}s_{i}^{A}{\otimes s}_{k}^{B}(\rho_{\text{\textit{in}%
}}^{\prime})s_{i}^{A\dagger}{\otimes s}_{k}^{B\dagger}.\label{sigma}%
\end{equation}
By taking the average of the operators (\ref{projectors}) over
$\rho _{\text{\textit{fin}}}$ we calculate the payoff function of
each player. The possible variations of the strategies are
introduced into the quantum noise terms ($\gamma_{A}$ and
$\gamma_{B}$). This process shows the quantum nature of the
game\footnote{In a open quantum system the operator $J^{\dagger}$
can be seen as a basis change of the payoff operators.}.

\begin{figure}[ht]
\centering 
\psfrag{deltaA}{$\gamma^{*}_A$} \psfrag{deltaB}{$\gamma^{*}_B$}
\psfrag{F}{}
\includegraphics[width=2.8in,angle=270]{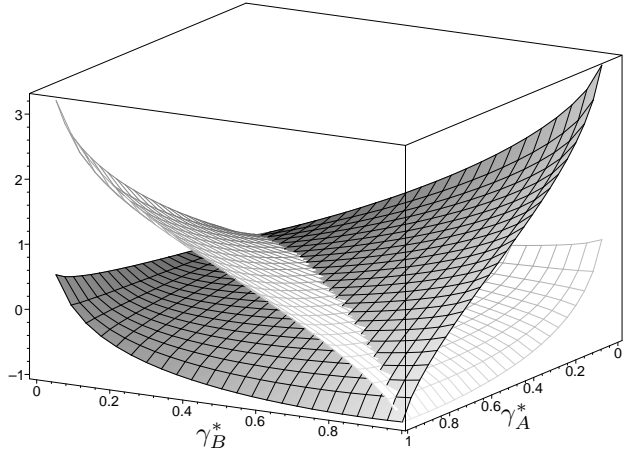}
\caption{\small{NE points for $(\gamma_{A},\gamma_{B}) =
(0.1,0.1)$
where $(\stackrel{\star}{\gamma_{A}},\stackrel{\star}{\gamma_{B}})_{\mathrm{MAX}}\approx(0.4,0.4)$.}}%
\label{fig:MaxQPDC_A}
\end{figure}

For a noise given by a QPDC, the channel $A$ plays against $B$
manipulating the states of the principal system through their set
of strategies $\mathcal{S}_{A}$ and $\mathcal{S}_{B}$. In this
way, we construct a game for the QPDC from $H_{TB}$. However, from
the rules of the superoperators (\ref{TB2 especificos}) and since
$J(\chi)|00\rangle=\cos(\chi/2)|00\rangle+i\sin(\chi
/2)|11\rangle$, the game model will not include the worst case in
a QPDC that occurs when the state is
$(|0\rangle+|1\rangle)\otimes(|0\rangle+|1\rangle)/2$. To outline
this problem, we apply the entanglement operator
$J_{PD}(\chi)\equiv J(\chi)(H\otimes H)$ instead of $J(\chi)$ in
EWL protocol. This new procedure does not modify the essential
structure of a quantum game and does not decrease the entanglement
degree\footnote{Since
$D=U(\pi,0)=|0\rangle\langle1|-|1\rangle\langle0|$, we have
$J_{PD}(\chi)|00\rangle=(-1)^{\chi/2\pi}(|00\rangle+|11\rangle)/2+(-1)^{-\chi/2\pi}(|01\rangle+|10\rangle)/2$
. Besides, $[J,s_{i}^{x}\otimes
s_{j}^{y}]=[J_{AD},s_{i}^{x}\otimes
s_{j}^{y}]=[J_{PD},s_{i}^{x}\otimes s_{j}^{y}]=0,$ $\forall i,j$
only when $s_{i}^{x}$ is equal to a generic classical strategy
$U_{i}(\theta_{x},0)$.}.

By applying (\ref{projectors}) over the final state (\ref{rho
final}), and considering a general $\chi,$ we obtain the
\textit{payoff function} of the player $A$:
\begin{equation}%
\begin{array}
[c]{c}%
\mathcal{F}_{A}(\gamma_{A},\gamma_{B})=I(s_{0}^{A},s_{0}^{B})\left(
\sqrt{1-\gamma_{{A}}}+\sqrt{1-\gamma_{{B}}}+1+\sqrt{1-\gamma_{{A}}}%
\sqrt{1-\gamma_{{B}}}\right)  /4\\
\\
+I(s_{0}^{A},s_{1}^{B})\left[  \sqrt{1-\gamma_{{A}}}-\sqrt{1-\gamma_{{B}}%
}+1-\sqrt{1-\gamma_{{A}}}\sqrt{1-\gamma_{{B}}}\right. \\
\\
\left.  +8\left(  \sqrt{1-\gamma_{{A}}}-\sqrt{1-\gamma_{{B}}}\right)
\cos\left(  \chi/2\right)  ^{2}+8\left(  \sqrt{1-\gamma_{{B}}}-\sqrt
{1-\gamma_{{A}}}\right)  \cos\left(  \chi/2\right)  ^{4}\right]  /4\\
\\
+I(s_{1}^{A},s_{0}^{B})\left[  \sqrt{1-\gamma_{{B}}}-\sqrt{1-\gamma_{{A}}%
}+1-\sqrt{1-\gamma_{{A}}}\sqrt{1-\gamma_{{B}}}\right. \\
\\
\left.  +8\left(  \sqrt{1-\gamma_{{A}}}-\sqrt{1-\gamma_{{B}}}\right)
\cos\left(  \chi/2\right)  ^{2}+8\left(  \sqrt{1-\gamma_{{B}}}-\sqrt
{1-\gamma_{{A}}}\right)  \cos\left(  \chi/2\right)  ^{4}\right]  /4\\
\\
+I(s_{1}^{A},s_{1}^{B})\left(  \sqrt{1-\gamma_{{A}}}\sqrt{1-\gamma_{{B}}%
}-\sqrt{1-\gamma_{{B}}}-\sqrt{1-\gamma_{{A}}}+1\right)  /4,
\end{array}
\label{Fitness de A JQD_PD}%
\end{equation}
where the joint information of each pair of strategies are
calculated from (\ref{inf conj e inf shannon}) and (\ref{prob
conj}). Similarly, the same calculation can be performed for the
player $B$.
\begin{figure}[ht]
\centering 
\psfrag{deltaA}{$\gamma^{*}_A$} \psfrag{deltaB}{$\gamma^{*}_B$}
\psfrag{F}{}
\includegraphics[width=2.8in,angle=270]{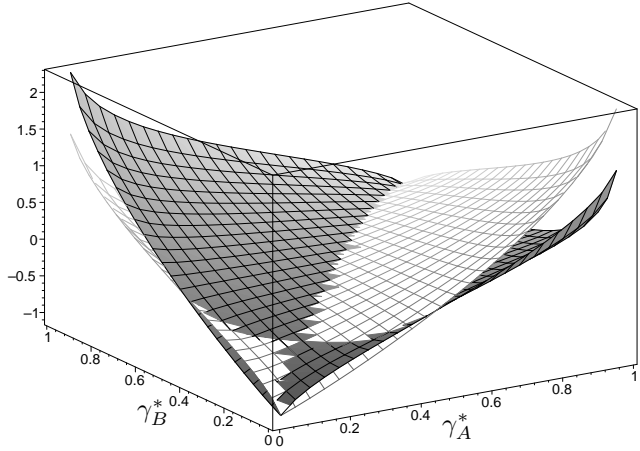}
\caption{\small{NE points for $(\gamma_{A},\gamma_{B}) =
(0.5,0.5)$
where $(\stackrel{\star}{\gamma_{A}},\stackrel{\star}{\gamma_{B}})_{\mathrm{MAX}}\approx(0.8,0.8)$.}}%
\label{fig:MaxQPDC_B}
\end{figure}

\begin{figure}[h]
\centering 
\psfrag{deltaA}{$\gamma^{*}_A$} \psfrag{deltaB}{$\gamma^{*}_B$}
\psfrag{F}{}
\includegraphics[width=2.8in,angle=270]{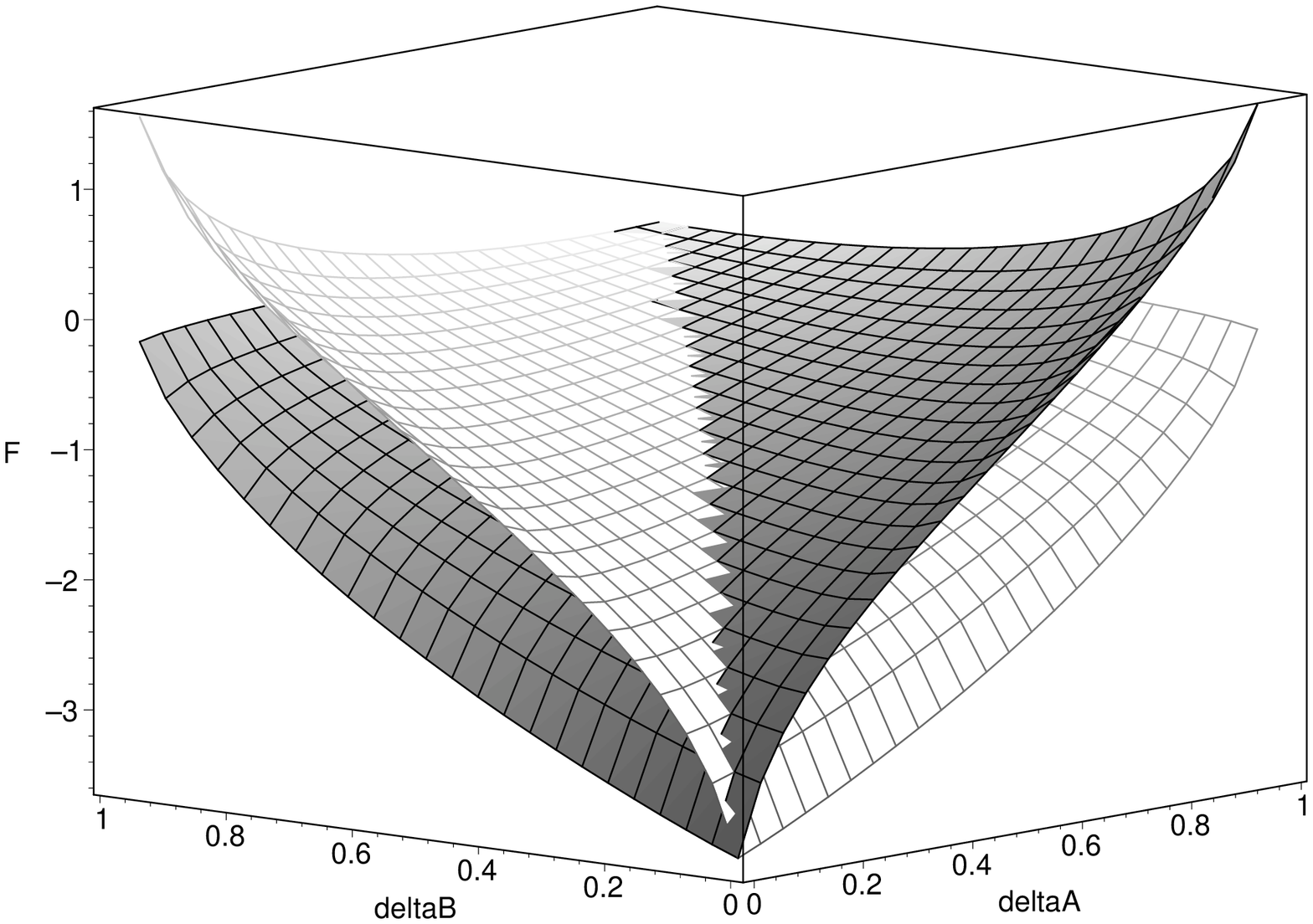}
\caption{\small{NE points for $(\gamma_{A},\gamma_{B}) = (1,1)$
where $(\stackrel{\star}{\gamma_{A}},\stackrel{\star}{\gamma_{B}})_{\mathrm{MAX}}\approx(1,1)$.}}%
\label{fig:MaxQPDC_C}
\end{figure}

\begin{figure}[h]
\centering 
\psfrag{deltaA}{$\gamma^{*}_A$} \psfrag{deltaB}{$\gamma^{*}_B$}
\psfrag{F}{}
\includegraphics[width=2.8in,angle=270]{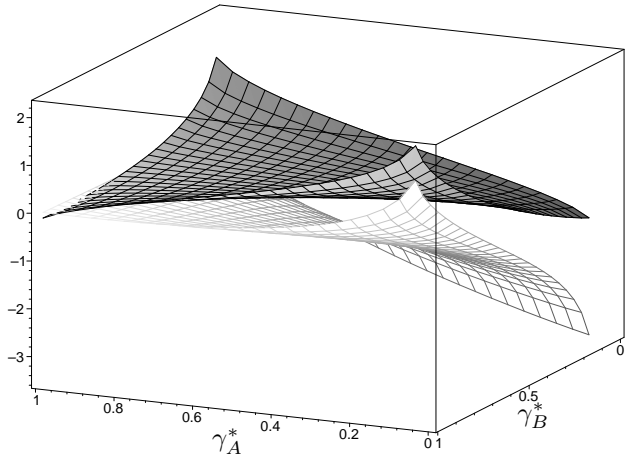}
\caption{\small{NE points for $(\gamma_{A},\gamma_{B}) = (1,0.5)$
where $(\stackrel{\star}{\gamma_{A}},\stackrel{\star}{\gamma_{B}})_{\mathrm{MAX}}\approx(1,0.8)$.}}%
\label{fig:MaxQPDC_D}
\end{figure}

To calculate the points of equilibrium we perform a numerical
analysis by using the Nash equilibrium concept (\ref{Nash A}) and
(\ref{Nash B}). By considering the initial state with a maximal
entanglement $J_{PD}(\pi/2)|00\rangle$ we calculate the pairs
$(\overset{\star}{\gamma}_{A},\overset{\star}{\gamma_{B}})$ for
four values of $(\gamma_{A},\gamma_{B})$ that generate the NE
points. Figures \ref{fig:MaxQPDC_A} to \ref{fig:MaxQPDC_D} show
the points where the vertical axis of the colorless surfaces
corresponds to the best equilibrium points (\ref{Nash A}) of the
player $A$; and the vertical axis of the colored surfaces
corresponds to the best equilibrium points (\ref{Nash B}) of the
player $B$. The points
$(\overset{\star}{\gamma}_{A},\overset{\star}{\gamma_{B}})$ in the
intersection of the surfaces that have positive values in the
vertical axis are Nash equilibrium points.

The colorless surface represents equation (\ref{Nash A}) and the
colored surface represents Equation (\ref{Nash B}).

The graphics show the correspondence between the equilibrium
points of the game and the physical stability of the system. By
physical stability we mean a larger time of coherence. But, a
larger time of coherence is related with a smallest probability of
noise which is associated with the smallest values of the
parameters of the strategies $\overset{\star}{\gamma}_{A,B}\sim0$.
It happens because the equilibrium points are directly associated
with the optimal points of information storage in the principal
system.

For small values of noise, $(\gamma_{A},\gamma_{B})=(0.1,0.1)$,
the game shows NE points
$(\stackrel{\star}{\gamma_{A}},\stackrel{\star}{\gamma_{B}})$ near
to zero (Figure \ref{fig:MaxQPDC_A}). For high values of noise,
$(\gamma_{A},\gamma_{B})=(1,1)$, the game shows NE points near to
one (Figure \ref{fig:MaxQPDC_C}). An interesting result is
achieved for $(\gamma _{A},\gamma_{B})=(0.5,0.5)$. The surfaces
that characterize the equilibrium points are very close (Figure
\ref{fig:MaxQPDC_B}). And a natural asymmetry of dominance of game
arise for different noise distribution, $\gamma_{A}\neq
\gamma_{B}$, $(\gamma_{A},\gamma _{B})=(1,0.5)$, without
eliminating the NE points (Figure \ref{fig:MaxQPDC_D}). Besides,
it is also possible to show the existence of NE points for
$(\gamma_{A},\gamma_{B})=(0.1,1);$
$(\gamma_{A},\gamma_{B})=(1,0.1)$ and for $(\gamma_{A},\gamma
_{B})=(0.5,0.1)$; $(\gamma_{A},\gamma_{B})=(0.1,0.5)$.

The main conclusion is that even under intense noise the system
may present physical stability points expressed by the game
equilibrium points.

\section{Discussion}

We discussed in section 3 that the Fröhlich model describes the
physical dynamics of a coupled biomolecule net. Inspired in this
dynamics, the Hamiltonian of Fröhlich has been used to model the
physical relation of the neuronal microtubules \cite{Tuszynski4}.
Several models claim that the microtubules process quantum
information in a sub-neuronal level. It would help to produce
certain mental functions such as the human consciousness
\cite{Faber2}--\cite{Hameroff1}.

It is our belief that many physical systems can be solved by
applying TGs to open quantum systems. Using the payoff criterion
described in this paper, we have linked the information storage
capacity to the physical stability points. Therefore, by using the
game models, we think it is possible a biophysical analysis that
shows the values of the relevant variables, such as the frequency
and temperature, in the equilibrium points. These models are able
to show whether there are or not equilibrium points close to the
physiological temperature and close to the protein frequencies.
This technique plays a fundamental role in describing any
information processing in biophysical systems.

All these issues deserve a detailed description. However, we
consider the TGs a very important tool to model, to calculate and
to interpret any quantum information processing in open systems.

\section{Conclusion}

In this paper we show how to construct a quantum game from
biophysical Hamiltonians. By using Kraus operators as quantum
strategies, we provide a possible game interpretation of open
quantum systems. In order to calculate the equilibrium points in
the Fröhlich quantum game we propose a new criterion for the
payoff matrix focusing on the information associated with the
principal system. By means of this criterion, the quantum game
model can be interpreted as a game between two different
environmental configurations which seek to minimize the noise
insertion into the principal system. However, to improve the
analysis of the equilibrium points, we have to incorporate some
physical conditions. To analyze all points in the QPDC game we
used the entanglement operator $J_{PD}(\chi)= J(\chi)(H\otimes H)$
instead of $J(\chi)$. Nevertheless, these extensions do not modify
the essence of a quantum game. The model shows physical
consistence and presents many points of stability, even under
intense noises ($ \gamma _{A,B}\sim1$).

Using this model the equilibrium of the game corresponds to the
points of Nash equilibrium which determine the maximum regimes of
information storage of the principal system. Therefore, the
description of a physical system as a quantum game gives us not
only a new interpretation but also a methodological alternative
for evaluating the information storage points in open systems.


\begin{thebibliography}{References}

\bibitem{Meyer} D. A. Meyer, Phys. Rev. Lett. 82 (1999) 1052.

\bibitem{Eisert1} J. Eisert, M. Wilkens, M. Lewenstein, Phys. Rev. Lett. 83 (1999) 3077.

\bibitem{Marinatto} L. Marinatto, T. Weber, Phys. Lett. A 272 (2000) 291.

\bibitem{Benjamim1} S. C. Benjamim, P. M. Hayden, Phys. Rev. A 64 (2001) 030301.

\bibitem{Du et al} J. Du, H. Li, X. Xu, M. Shi, J. Wu, X. Zhou, R. Han, J. Phys. A 36 (2003) 6551.

\bibitem{Abbott1} A. P. Flitney, D. Abbott, Fluct. Noise Lett. 4 (2002) R175.

\bibitem{Chen et al} L. K. Chen, Huiling Ang, D. Kiang, L. C.
Kwek, C. F. Lo, Phys. Lett. A 316 (2003) 317.

\bibitem{Chuang} M. A. Nielsen, I. L. Chuang, Quantum
Computing and Quantum Information, Univ. Press, Cambridge (2000).

\bibitem{Frohlich1} H. Fröhlich, Int. J. Quantum Chem. 2 (1968) 641.

\bibitem{Pokorny} J. Pokorný, T-M. Wu, Biophysical
Aspects of Coherence and Biological Order, Springer (1998).

\bibitem{Hofbauer} J. Hofbauer, K. Sigmund, Evolutionary Games
and Population Dynamics, Univ. Press, Cambridge (1998).

\bibitem{Tuszynski4} J. A. Tuszy\'{n}ski, J. A. Brown, P. Hawrylak,
The Royal Society 356 (1998) 1897.

\bibitem{Faber2} J. Faber, R. Portugal, L. P. Rosa, BioSystems 83 (2006) 1.

\bibitem{Faber1} J. Faber, L. P. Rosa, Phys. Rev. E 70 (2004) 031902.

\bibitem{Hameroff1} S. Hagan, S. R. Hameroff, J. A Tuszynski,
Phys. Rev. E 65 (2002) 061901.


\end{thebibliography}
\end{document}